\documentclass[journal,twoside,web]{ieeecolor}
\usepackage{generic}
\usepackage{amsmath,amssymb,amsfonts}
\usepackage{graphicx}
\usepackage{textcomp}
\usepackage{epstopdf}
\usepackage{siunitx}
\usepackage{amssymb}
\usepackage{subfig}
\usepackage{cite}
\usepackage{ mathrsfs }
\def\BibTeX{{\rm B\kern-.05em{\sc i\kern-.025em b}\kern-.08em
		T\kern-.1667em\lower.7ex\hbox{E}\kern-.125emX}}
\begin{document}
	\title{Ultra-thin Junctionless Nanowire FET Model, Including 2D Quantum Confinements} 
	\author{Danial Shafizade, Majid Shalchian, and Farzan Jazaeri
		\vspace{-0.6cm}
		\thanks{
		Danial Shafizade and Majid Shalchian are with the Electrical Engineering Department, Amirkabir University of Technology, 424 Hafez Ave, Tehran, Iran, 15875-4413. Farzan Jazaeri is with the Integrated Circuits Laboratory (ICLAB) of the Ecole Polytechnique F\'{e}d\'{e}rale de Lausanne, Switzerland.}}
	\maketitle
	\begin{abstract}
		In this paper, we develop an explicit model to predict the DC electrical behavior in ultra-thin surrounding gate junctionless nanowire FET. The proposed model takes into account 2D electrical and geometrical confinements of carrier charge density within few discrete sub-bands. Combining a parabolic approximation of the Poisson equation, first order perturbation theory for the Schrödinger subband energy eigenvalues, and Fermi-Dirac statistics for the confined carrier density leads to an explicit solution of the DC characteristic in ultra-thin junctionless devices. Validity of the model has been verified  with technology computer-aided design simulations. The results confirms its validity for all regions of operation, i.e., from deep depletion to accumulation and from linear to saturation. This represents an essential step toward analysis of circuits based on junctionless nanowire devices.
	\end{abstract}
\section{Introduction}
\IEEEPARstart{J}{unctionless} (JL) Silicon Nanowire FETs used heavily doped channel to relax several critical fabrication steps and to obviate formation of source/channel and drain/channel junctions, which degrade the performance of short channel devices \cite{Nature}. To deploy full advantages of Gate-All-Around (GAA) JL devices, several compact model have been proposed so far. Particularly in \cite{Book}, \cite{5872019}, we developed a charge-based model, using Poisson-Boltzmann equations to model planar and cylindrical junctionless field effect transistors (JLFETs). This model, works for JLFETs with the channel thicknesses down to 10 nm and the device characteristics are well-captured by the proposed model. Nevertheles, this model neglects charge quantization into discrete subband, which is not a valid assumption as channel size decrease below 10 nm. As demonstrated in Fig. 1, the classic charge-based model in \cite{5872019} fails to predict the charge density in the channel thickness of $T_{SC}=4$ nm.

 To accurately capture the quantization effect, a coupled Poisson-Schr\"{o}dinger equation (PS) needs to be solved self-consistently.This is a time-consuming calculation method and computationally not suitable for compact modeling purposes.
 Another work \cite{314027} proposed to incorporate the influence of charge quantization as a shift in the DC characteristics. However, due to the coupling between Poisson and Schr\"{o}dinger equations, this shift varies with the gate to source voltage and therefore this method is not valid in all regions of operation. Lately, taking into account 1D quantum confinement, we developed an analytical approach for ultra-thin junctionless double-gate FETs, which introduces a zero order approximation in the wave-functions and a first order correction for the confined energies in all the regions of operation \cite{8424086}. In this context, extending the proposed approach  we  derive an explicit model for junctionless GAA nanowire FET with ultra-thin rectangular cross section. 
\begin{figure}[!t]
	\centering
	\includegraphics[ height=4.5cm, width=8cm]{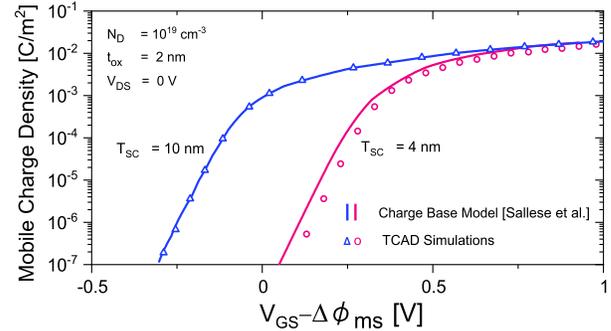}
	\caption{logarithmic drain current versus the effective gate voltage for
		different $V_{DS}$. Solid lines: classic charge-based model \cite{5872019}. symbols: TCAD simulation based on the classic Boltzmann statistics. Dashed lines: TCAD simulations based on Poisson–Schrödinger. ${t}_{ox}=2$ nm, ${T}_{sc}=4,10$ nm, ${N}_{D}={10}^{19}$ cm$^{-3}$, $V_{ch}=0$ V}
	\label{fig1}
	\vspace{-0.6cm}
\end{figure}
\section{Device structure and Model Derivation}
We assume an \textit{n}-type junctionless nanowire GAA with a rectangular  cross section perpendicular to $ Z $ direction, shown in Fig. 2(a) where both the channel thickness (along vertical direction, i.e. $ Y $) and the channel width (along lateral direction, i.e. $ X $) are equal to ${T}_{sc}$ and the gate oxide thickness is ${t}_{ox}$ all around the silicon channel. The silicon channel is doped uniformly with a high level of donor concentration i.e.  ${N}_{D}$ and other physical parameters are given in Table I, used in the model derivations
and TCAD simulations. The principle of operation for junctionless devices is different from the regular junction-based double-gate device as the current flows through the volume instead of Si$-$SiO$_2$ interfaces \cite{8357582}. A junctionless nanowire FET transistor has three modes of operation \cite{Book}: While the gate to source potential, i.e. ${V}_{GS}$, is below the flat-band condition at the source terminal, i.e. ${V}_{FB,S}$ device operates in depletion mode. Whereas ${V}_{GS}$ is above the flat-band condition at the drain terminal, i.e. ${V}_{FB,D}$ device is in the accumulation region. A peculiar situation inherent to junctionless FETs is when depletion and accumulation coexist inside the channel, what is called hybrid channel state \cite{Book}. This happens when part of the channel near the source is in depletion whereas it turns into accumulation near the drain. 

In the following section, applying the proposed treatment in \cite{6403540} for 2D carrier confinement, the electrostatic potential profiles are accurately predicted in junctionless GAA nanowire FET for $3$ to $5$ nm silicon thicknesses.
\vspace{-0.5cm}
\begin{center}
	\begin{table}
		\caption{Physical Parameters of GAA JL Nanowire FET used in the TCAD simulations and Model derivations}
		\renewcommand{\arraystretch}{1.5}
		\begin{tabular}{p{4cm}p{1.3cm}p{2.3cm}}
			\hline
			\textbf{Parameter} &\textbf{Symbol}&\textbf{Value}  \\
						\hline
						\hline
			\textbf{Channel Doping} &${N}_{D}  $&$10^{19} $ cm$ ^{-3} $  \\
\textbf{Channel Thickness}  &${T}_{sc} $& $3-5$ nm  \\
\textbf{Oxide Thickness}  &${t}_{ox} $& $1$ nm  \\
\textbf{Channel Width}&${L}_{G} $&$1.2$ $\mu$m  \\
\textbf{Permittivity in Vacuum }&$\varepsilon_{o}$ &$8.85{\times}10^{-12}$ F/m  \\
\textbf{Silicon Permittivity}&$\varepsilon_{si}$&$11.7\varepsilon_{o}$   \\
\textbf{Silicon Oxide Permittivity}&$\varepsilon_{ox}$&$3.9\varepsilon_{o}$   \\
\textbf{longitudinal  Effective mass }&$m_{l}$ &$0.916m_{o}$  \\
\textbf{Transverse Effective mass}&$m_{t}$ &$0.19m_{o}$  \\
\textbf{Silicon Band Gap}&$E_{g}$ &$1.12$ eV  \\
\textbf{Gate Work Function}&$\Delta\phi_{m}$ &$4.8$ eV  \\
\textbf{Conduction Band effective DoS}&$N_{c}$ & $2.8{\times}10^{19}$ cm$^{-3}$  \\
\textbf{Valence Band Effective DoS}&$N_{v}$ &$1.04{\times}10^{19}$ cm$^{-3}$  \\
\textbf{Silicon Intrinsic Carrier Density}&$n_{i}$ &$1.45{\times}10^{10}$ cm$^{-3}$    \\
\textbf{Temperature}&$T$ &$300^{\circ}$ K \\
\textbf{Mobility}&$\mu_{n}$ &$90.5 $ cm$^{-3}$/V.S  \\
\textbf{Thermal Voltage}&$v_{t}$ & $0.025$ V  \\
			\hline
		\end{tabular}
	\end{table}
\end{center}
\section{Electrostatics in Ultrathin Junctionless GAA nanowire FETs}
The 3D electrostatic potential distribution obtained from the TCAD simulation results is shown in Fig. 2(b) for $V_{GS}=0.3$ V and $ 4 $ nm channel thickness doped at $10^{19} $cm$^{-3}$. 
Here, we propose to follow the same approach in \cite{Ferrier_2006}, and to assume a parabolic approximation for the 2D potential distribution $ \psi(x,y) $ in lateral and vertical directions in an ultra-thin junctionless GAA (for both depletion and accumulation regions). Therefore, the parabolic approximation can be expressed as:
\begin{equation} \label{1}
\!\psi(x,y)\!=\!{\psi}_{s}\!-\!2({\psi} _{s}\!-\!{\psi}_{c})\left[\frac{x}{{{T}_{sc}}}\left(\!1\!-\!\frac{x}{{{T}_{sc}}}\!\right)\!\!+\!\!\frac{y}{{{T}_{sc}}}\left(\!1\!-\!\frac{y}{{{T}_{sc}}}\!\right)\right]
\end{equation}
as shown in Fig. 2(c)  ${\psi }_{c}$ and ${\psi }_{s}$ are respectively the electrostatic potentials at the center and corners of the rectangular channel. 
In addition, the boundary conditions arising from the continuity of the displacement vectors at the four silicon-insulator interfaces of the square cross section give the total semiconductor charge per unit length:
${Q}_{sc}=4{{\varepsilon }_{si}}{{E}_{sc}}{{T}_{sc}}$, where ${E}_{sc}$ is the magnitude of the electric field, obtained from (1). Therefore,  
\begin{equation} \label{2}
{Q}_{sc}=-4{{\varepsilon }_{si}}{T}_{sc}\frac{\partial \psi (x,y)}{\partial y}\bigg| _{int} =8{{\varepsilon }_{si}}\Delta V,
\end{equation}
where $\Delta V={\psi} _{s}\!-\!{\psi}_{c}$ is directly proportional to  ${Q}_{sc}$. The total charge density is shared among four silicon-insulator capacitors around the channel and an average voltage drop over the gate oxide barrier is defined by $\bar{V}_{ox} = {Q}_{sc}/4{C}_{ox}{T}_{sc}$ where ${C}_{ox}\!=\!{{{\varepsilon }_{ox}}}/{{{t}_{ox}}}$  corresponds to the gate oxide capacitance per unit area. On the other hand, $V_{ox}$ varies along the channel interface, leading to 
\begin{align}
& {{V}_{ox}(y)}+\psi (0,y)={{V}_{ox}(x)}+\psi (x,0)={{V}_{GS}}-{{\Delta\phi }_{ms}}.
\end{align}

\begin{figure}[!t]
	\centering
	\includegraphics[ height=8cm, width=8.5cm]{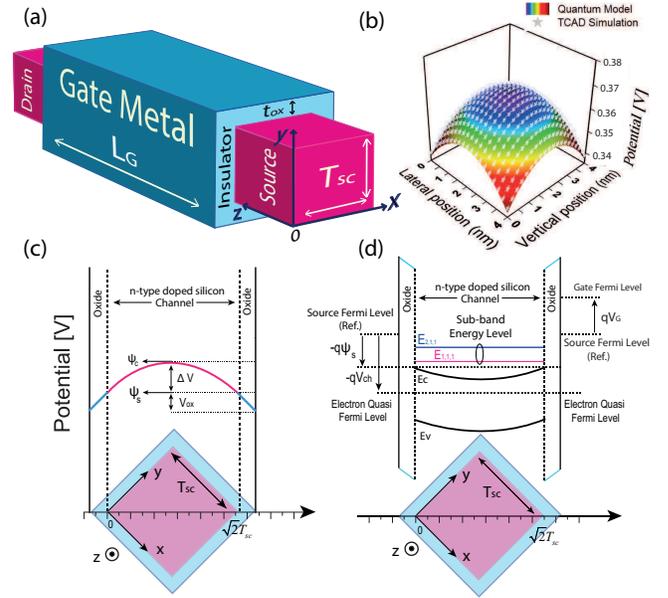}
	\caption{(a) 3D view of the junctionless GAA nanowire FET (b) Potential distribution across the channel cross section, for $V_{GS}=0.3$ V and ${T}_{sc}=4$ nm predicted by the model and validated by TCAD simulation results (c) Potential, and (d) Energy band-diagram across the diagonal of the square cross section. }
	\label{fig2}
	\vspace{-0.2cm}
\end{figure}
To account for this and to obtain ${\psi }_{s}$, we integrate (3) around each corner of the rectangle. For instance, once integrating (3) at $x=0$ and $y=0$ from $y=0$ to $y={T}_{sc}/{2}$ over $ y $-axis and from $x=0$ to $x={T}_{sc}/{2}$ over the $x$-axis leads to the following expression for $ \bar V_{ox} $:  
\begin{equation}
\begin {split}
\!\!\!\!\bar V_{ox}\!=\!{V_{GS}}\!-\!{{\Delta\phi }_{ms}}\!-\!\frac{1}{{T}_{sc}}\left(\!\!\int\limits_{0}^{\frac{T_{sc}}{2}}   \psi(0,y)dy\!+\!\int\limits_{0}^{\frac{T_{sc}}{2}}\psi (x,0)dx\right)\!\!
\end {split}
\end{equation}
By substituting the terms of ${\psi }(x,0)$ and ${\psi }(0,y)$ obtained from (1) into (4), we get $\bar V_{ox}$, leading to a key relation between ${\psi }_{s}$, ${V}_{GS}$ and $\Delta V$:
\begin{align}
&{{\psi }_{s}}={{V}_{GS}}-{{\Delta\phi }_{ms}}+\Delta V\left(\frac{1}{3}-2\frac{{{C}_{si}}}{{{C}_{ox}}}\right)
\end{align}
 \begin{figure}[!t]
	\centerline{\includegraphics[ height=8cm, width=8cm]{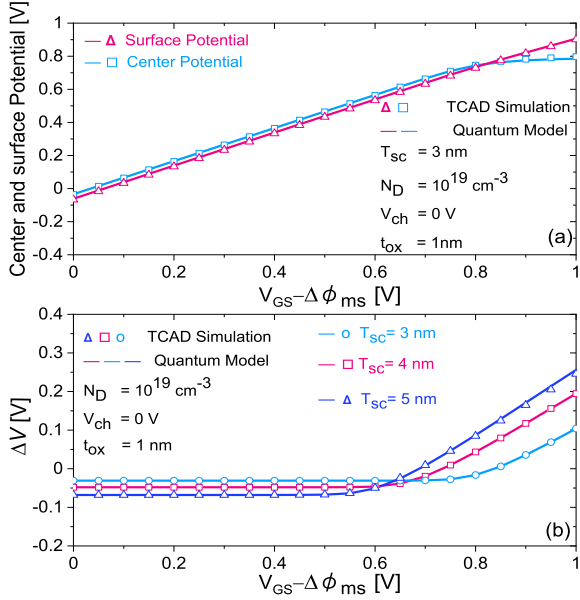}}
	\caption{(a) Surface potential ${\psi }_{s}$ and Center Potential ${\psi }_{c}$ as functions of electrostatic potential for ${T}_{sc}=3$ nm  (b)  ${\Delta V}$ as a function of gate voltage for channel thicknesses of $3$, $4$ and $5$ nm Solid lines: proposed model. symbols: TCAD simulation.}
	\label{fig3}
	\vspace{-0.2cm}
\end{figure}
where ${C}_{si}\!=\!{{{\varepsilon }_{si}}}/{{{T}_{sc}}}$  corresponds to semiconductor capacitance per unit area. Next, relying on the 1st order perturbation theory, we formulate the charge quantization in discrete sub-bands. Due to two-dimensional (2D) geometrical and electrical confinements in ultra-thin nanowire, mobile charges are confined in discrete sub-bands which are the solutions of the Schr\"{o}dinger equation in the $x$, $y$ plane. An analytical solution can be obtained by assuming an ideal infinite potential well boundary condition at Si-SiO$_2$ interface along with the 1st order perturbation potential inside the channel. For zero order solution, sub-bands energies and wave-functions are given by:
\begin{align}
{{E}_{k,{{n}_{x}},{{n}_{y}}}}=\frac{{{\hbar }^{2}}{{\pi }^{2}}{{n}^2_{x}}}{2m_{kx}^{*}{{T}^2_{sc}}}+\frac{{{\hbar }^{2}}{{\pi }^{2}}{{n}^2_{y}}}{2m_{ky}^{*}{{T}^2_{sc}}},
\end{align}
\begin{align}
\Psi (x,y)=\sqrt{\frac{2}{{{T}_{sc}}}}\sin \left(\frac{\pi {{n}_{x}}x}{{{T}_{sc}}}\right)+\sqrt{\frac{2}{{{T}_{sc}}}}\sin \left(\frac{\pi {{n}_{y}}y}{{{T}_{sc}}}\right),
\end{align}
where $\Psi$ is the 2D wave function, and $\psi (x,y)$ is the 2D electrostatic potential in the channel and ${m}_{kx}^{*}$ and ${m}_{ky}^{*}$ are electron effective mass which depends on silicon orientation. In case of $<\!100\!>$ orientation, three different series must be considered to properly account for 6 valleys of the effective mass ellipsoid. 
$({m}_{kx}^{*},{m}_{ky}^{*})={(m_l,m_t),(m_t,m_l),(m_t,m_t)}$, $m_l=0.916m_0$ and $m_t=0.19m_0$ are the longitude and transversal effective mass of electron in silicon, respectively. ${n}_{x}$ and ${n}_{y}$  are the quantum numbers for $x$ and $y$ directions that starting from $({n}_{x},{n}_{y})=(1,1)$ which is the first sub-band. For the first sub-band, due to the degeneracy of 6 valleys in two groups, only two degenerate energy levels exist. The first order correction to the energy eigenvalues is computed using the time-independent perturbation theory:
\begin{align}
E_{{{n}_{x},{n}_{y}}}^{p}=q\int\limits_{0}^{{T}_{sc}}\int\limits_{0}^{{T}_{sc}}\left\{{\left[ \Psi (x,y) \right]^{*}{\psi }(x,y){{\left[ \Psi (x,y) \right]}}}\right\}dxdy.
\end{align}
Replacing ${\psi }_{s}$ obtained  from (5) into (1), and the resulting $\psi (x,y)$ in (8), the integral can be solve analytically and the first order correction of sub-band energies is obtained as a function of $\Delta V$, given by
\begin{align}
E_{{{n}_{y}},{{n}_{x}}}^{p}=q\Delta V\left(\frac{2}{3}+\frac{1}{n_{x}^{2}{{\pi }^{2}}}+\frac{1}{n_{y}^{2}{{\pi }^{2}}}\right).
\end{align}
	\begin{figure}[!t]
	\centering 
	\includegraphics[height=8cm, width=8cm]{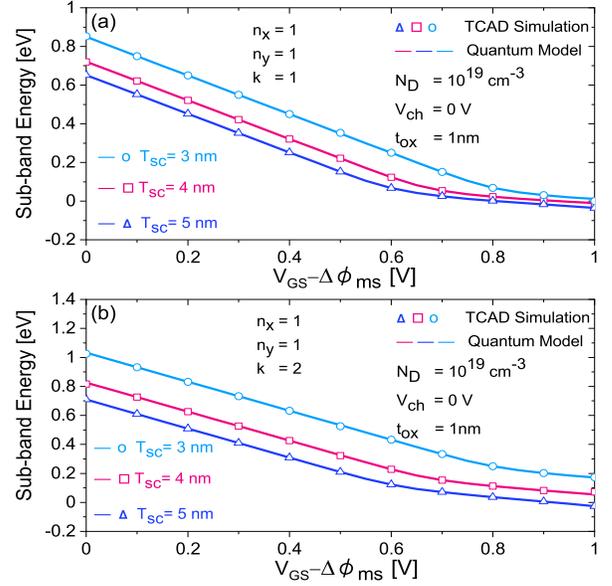}
	\caption{1st sub-band energy levels from the reference of electron quasi-fermi level at source as a function of gate voltage for channel thicknesses of $3$, $4$ and $5$ nm.(a) first degenerate level $E_{1,1,1}^{T}-q{\psi }_{s}$ (b) second degenerate level $E_{2,1,1}^{T}-q{\psi }_{s}$ calculated from analytical  model (Solid lines) and verified by TCAD simulation (symbols).}
	\label{fig4}
	\vspace{-0.3cm}
\end{figure}
\begin{figure}[!t]
	\centering
	\includegraphics[height=8cm, width=8cm]{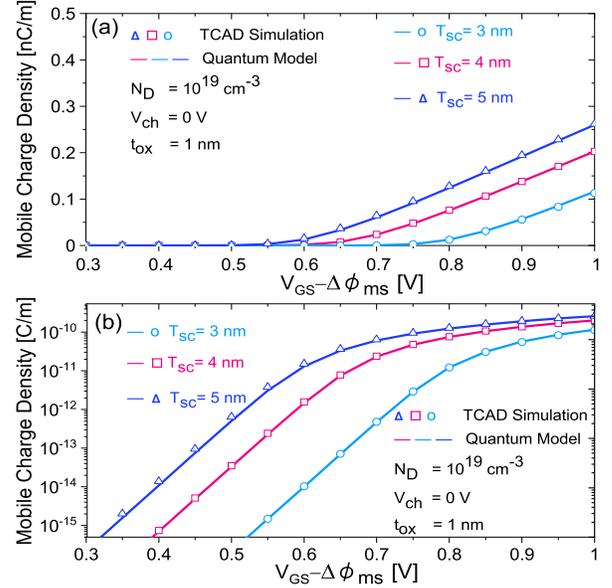}
	\caption{Mobile charge density with respect to the effective gate to source voltage for different channel thicknesses of $3, 4,$ and $5$ nm (a) linear scale and (b) semi-log scales. Solid lines: proposed model. symbols: TCAD simulation.}
	\label{fig5}
\end{figure}
\begin{figure}[!t]
	\centerline{\includegraphics[height=8cm, width=8cm]{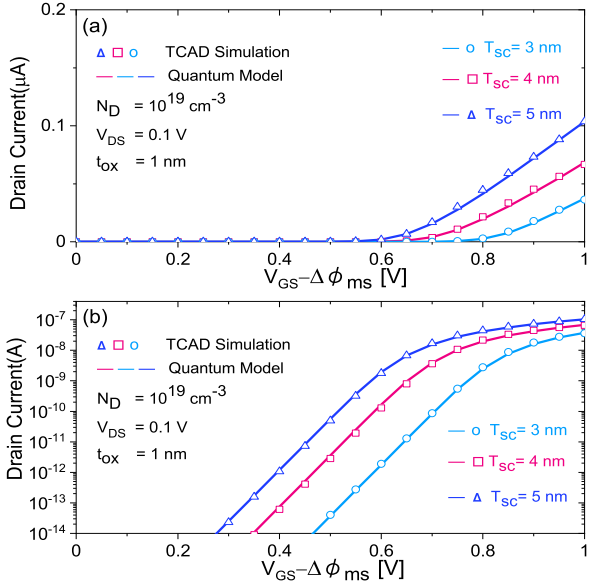}}
	\caption{Drain current for channel thickness of $3, 4,$ and $5$ nm versus $V_{GS}$  for ${V}_{DS}=0.1 $ V (a) linear and (b) semi-log scales. Solid lines: proposed model. symbols: TCAD simulation.}
	\label{fig6}
	\vspace{-0.5cm}
\end{figure} 
The total energy arising from geometrical and electrical confinements is the summation of (6) and (9), given by
\begin{align}
{E_{k,{{n}_{x}},{{n}_{y}}}^{T}}={{E}_{k,{{n}_{x}},{{n}_{y}}}}+E_{{{n}_{y}},{{n}_{x}}}^{p}.
\end{align}
The term of $E_{k,{{n}_{x}},{{n}_{y}}}^{T}$ is measured from the reference of the surface electron energy $-q{\psi }_{s}$ as shown in Fig. 2(d). Total semiconductor charge density per unit length in the nanowire is the sum of fixed charges and mobile charges:
\begin{align}
{{Q}_{sc}}={{Q}_{m}}+{{Q}_{fix}},
\end{align}
where ${{Q}_{fix}}=q{{N}_{D}}T_{sc}^{2}$ represents the fix charge density per unit length and  ${{Q}_{m}}$ is corresponding to the mobile charge density which can be written by using Fermi integral in order of $-\frac{1}{2}$  as function of $\Delta V$:
\begin{align}
{{Q}_{m}}=-\sum\limits_{k=1}^{2}{qDO{{S}_{k}}{{F}_{-\frac{1}{2}}}}\left( -\frac{\eta }{{{K}_{B}}T} \right),
\end{align}
where $\eta$ is the distance between the sub-band level and electron quasi-Fermi level in the channel, i.e $q{V}_{ch}$, the latter  varies from $0$  at source to $-q{V}_{DS}$ at the drain, therefore,
\begin{align}
\eta=E_{k,{{n}_{x}},{{n}_{y}}}^{T}-{q{\psi}_{s}}+{q{V}_{ch}} ,
 \end{align}
and $  
Do{{S}_{k}}={g}_{k}\sqrt{\frac{2m_{k}^{*}{{K}_{B}}T}{{{\hbar }^{2}}{{\pi }^{2}}}}
$
 is one dimensional (1D) effective density of states in nanowire cross section and ${{g}_{k}}$  is the degeneracy factor, $(g_1, g_2)=(4,2)$ for the first sub-band. For a given bias point, (5) and (10) describe the surface potential ${\psi }_{s}$ and sub-band energy, $E_{k,{{n}_{x}},{{n}_{y}}}^{T}$, as a function of $\Delta V$. Replacing $Q_{sc}$ from (2) and $Q_m$ from (12) into (11), a single charge-based equation is solved semi-analytically to obtain the electron energy level, potential profile, and charge density. Fig. 3 shows the surface and center potentials and $\Delta V$ as a function of ${V}_{GS}$. In depletion mode (for small value of ${V}_{GS}-\Delta\phi_{ms}$), the mobile charge density is almost negligible, therefore $\Delta V$ is equal to  $-qN_DT_{sc}^2/(8\varepsilon_{si}) $, and the channel center potential becomes higher than the surface potential, causing an electrical confinement of the carriers wave-function into discrete 2D sub-bands, which enhances the geometrical 2D confinement in the channel. By increasing (${V}_{GS}$ toward the $V_{FB}$ this electrical confinement is relaxed. Fig. 4(a) and 4(b) shows degenerate energy levels for the first sub-band $({n}_{x},{n}_{y})=(1,1)$ versus ${V}_{GS}$  for various channel thickness ($T_{sc}=3,4,5$ nm). This figure clearly shows that the electrical confinement is relaxed by increasing the gate voltage up to the flat-band potential. Fig. 5 represents the mobile charge density versus the effective gate voltage for different channel thicknesses i.e. $ T_{sc}=3-5$ nm calculated from the proposed model and verified by two-dimensional TCAD simulations results.

\begin{figure}[!t]
	\centerline{\includegraphics[height=8cm, width=8cm]{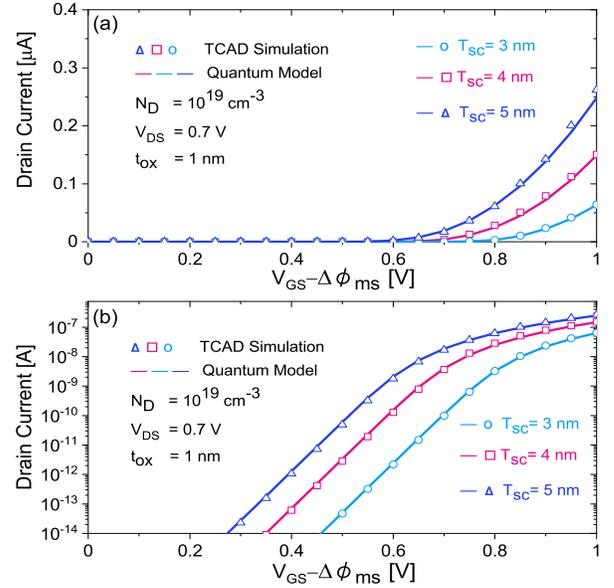}}
	\caption{Drain current for channel thickness of $3,4,5$ nm as a function of $V_{GS}$  for ${V}_{DS}=0.7 $ V (a) linear scale and (b) semi-log scale. Solid lines: proposed model. symbols: TCAD simulation.}
		\label{fig7}
	\vspace{-0.5cm}
\end{figure} 
\section{Derivation of the Drain Current}
As in \cite{27041}, relying on the drift-diffusion transportation model, we propose to calculate the drain current, given by:
\begin{align}
{{I}_{DS}}=-\frac{\mu  q}{L_G}\int\limits_{{{\eta}_{S}}}^{{{\eta}_{D}}}{{{Q}_{m}}dV_{ch}}
\end{align}
where $ \mu $ is the free carrier mobility. We intentionally assume a constant mobility to focus on the essentials of electrostatics. Here, ${\eta}_{S}=E_{k,{{n}_{x}},{{n}_{y}}}^{T}-q{\psi}_{s}$ and ${\eta}_{D}=E_{k,{{n}_{x}},{{n}_{y}}}^{T}+q{V}_{DS}-q{\psi}_{s}$ represent the differences between the sub-band and quasi-Fermi levels at the source and drain sides. Replacing (12) into (14), and applying the chain rule for d${V}_{ch}$, we can write:
\vspace{-0.2cm}
\begin{equation}
{{I}_{DS}}=\frac{\mu }{L_G}\sum\limits_{K=1}^{2}{qDo{{S}_{K}}}\int\limits_{{\eta }_{S}}^{{\eta }_{D}}{{{{F}_{ -\frac{1}{2}}}}\left(-\frac{\eta}{{{K}_{B}T}}\right)} \frac{dV_{ch}}{d\eta}d\eta.
\end{equation}
The term of ${dV_{ch}}/{d\eta}$ is obtained from the definition of $\eta$ as:
\begin{equation}
\frac{d\eta}{dV_{ch}}=q\left[ 1+\frac{d{{\psi }_{s}}}{d{{V}_{ch}}}\left( \frac{dE_{k,{{n}_{x}},{{n}_{y}}}^{T}}{d{{\psi }_{s}}}-1 \right) \right].	
\end{equation}
Replacing $\Delta V$ from (2) into (5), we can write: ${{V}_{GS}}-{\Delta{\phi}_{ms}}-{{\psi }_{s}}=-{{Q}_{sc}}/{{\beta\varepsilon}_{ox}}$, where $\beta={24\varepsilon_{si}}/({\varepsilon_{ox}-6\varepsilon_{si}})$. Differentiating both sides of this relationship with respect to the $ {V}_{ch} $ we get:
\begin{equation}
\frac{d{{\psi }_{s}}}{d{V}_{ch}}=\frac{1}{{\beta\varepsilon_{ox}}}\frac{d{{Q}_{m}}}{d{V}_{ch}}.
\end{equation}
Next, applying the chain rule,  ${d{E_{k,{{n}_{x}},{{n}_{y}}}^{T}}}/{d{{\psi }_{s}}}$ is obtained as follows
\begin{equation}
\frac{d{E_{k,{{n}_{x}},{{n}_{y}}}^{T}}}{d{{\psi }_{s}}}=\frac{d{E_{k,{{n}_{x}},{{n}_{y}}}^{T}}}{d\Delta V}\frac{d\Delta V}{d{{\psi }_{s}}}.
\end{equation}
Using (5) and (10) we can express (18) by
\begin{equation}
\frac{d{E_{k,{{n}_{x}},{{n}_{y}}}^{T}}}{d{{\psi }_{s}}}=2\left(\frac{1}{3}+\frac{1}{{{n}^{2}}{{\pi }^{2}}}\right)\left(\frac{1}{3}-2\frac{{{C}_{si}}}{{{C}_{ox}}}\right)^{-1}=a.
\end{equation}
Now we can rewrite (16) based on (17) and (19) as follows
\begin{equation}
\frac{d\eta}{dV_{ch}}=q\left[ 1+\left( a-1 \right)\left( \frac{1}{\beta\varepsilon_{ox}}\frac{d{{Q}_{m}}}{d{{V}_{ch}}} \right) \right].
\end{equation}
Replacing derivative of $Q_m$ with respect to ${V}_{ch}$ from (12) into (20) we have
\begin{align}
\frac{d
	\eta}{d{V}_{ch}}=q\left[ 1+\alpha\sum\limits_{k=1}^{2}q{Do{{S}_{k}}\frac{d{{F}_{-\frac{1}{2}}}(-\frac{\eta }{{{K}_{B}}T})}{d\eta }\frac{d\eta }{d{{V}_{ch}}}} \right],
\end{align}
where $\alpha= (a-1)/\beta\varepsilon_{ox}$ taking into account (19) depends on channel and oxide thickness, therefore ${dV}_{ch}/{d\eta}$  is given by
\begin{equation}
\frac{d{V}_{ch}}{d\eta}=\left[ \frac{1}{q}-\alpha \sum\limits_{k=1}^{2}{qDo{{S}_{k}}\frac{d{{F}_{-\frac{1}{2}}}(-\frac{\eta }{{{K}_{B}}T})}{d\eta }} \right].
\end{equation}
Applying (22) into (15) we can obtain the drain current
\begin{equation}
\begin{split}
&{{I}_{DS}}=\frac{\mu }{L_G}\sum\limits_{K=1}^{2}q{Do{{S}_{K}}}\int\limits_{{\eta }_{S}}^{{\eta }_{D}}{{{{F}_{-\frac{1}{2}}}}\left(-\frac{\eta}{{{K}_{B}T}}\right)}\\
&\times d\eta \Bigg [ \frac{1}{q}-\alpha \sum\limits_{k=1}^{2}{qDo{{S}_{k}}\frac{d{{F}_{-\frac{1}{2}}}(-\frac{\eta }{{{K}_{B}}T})}{d\eta}} \Bigg ].
\end{split}
\end{equation}

Integrating (23) from source to drain we get
\begin{equation}
\begin{split}
&{{I}_{DS}}=\frac{\mu {{K}_{B}}T}{{{L}_{G}}}\sum\limits_{k=1}^{2}{Do{{S}_{k}}}\left[ {{F}_{\frac{1}{2}}}(-\frac{\eta }{{{K}_{B}}T}) \right]_{{{\eta }_{S}}}^{{{\eta }_{D}}}\\
&-\frac{\mu\alpha }{2{{L}_{G}}}\sum\limits_{k=1}^{2}{{{q}^{2}}DoS_{k}^{2}}\left[ \Bigg({{F}_{-\frac{1}{2}}}(-\frac{\eta }{{{K}_{B}}T}) \Bigg)^{2}\right]_{{{\eta }_{S}}}^{{{\eta }_{D}}},\\
\end{split}
\end{equation}
Fig. 6 and Fig. 7 show the drain current versus the effective gate voltage for low ${V}_{DS}$ ($=0.1$ V) and high ${V}_{DS}$ ( $=0.7$ V) respectively, the length of the channel is ${L}_{G}=1.2$ $\mu $m to avoid any short and narrow channel effects. The result predicted by the model shows excellent agreement with TCAD simulations, in both linear and saturation conditions for both charge depletion and accumulation modes. This confirms validation of the model under every operation regions. 

Finally, it is worth mentioning that to derive the proposed analytical model, few approximations have been made. First, we assumed infinite quantum well with the first order perturbation theory for sub-band energy estimation and we estimated  the potential distribution in channel cross-section by using the parabolic approximation,  besides, only the contribution of the first sub-band has been considered in the drain current derivation. These approximations are valid for ultra-thin silicon channel. Nevertheless, as $T_{sc}$ increases above $6$ nm, the mobile charge density is distributed non-uniformly near the boundaries of the silicon channel and forms a triangular quantum well, which causes a deviation from the pre-assumed parabolic potential approximation and also affects sub-band energies. Moreover, the contribution of higher sub-bands should not be neglected due to the relaxation of geometrical confinement.

\section{Conclusion}
In this paper, we derived an analytical model to calculate the two dimensional electrostatic potential distribution in ultra-thin junctionless GAA nanowire FET with channel thickness in the range of $ 3$ to $6 $ nm.  Relying on the drift-diffusion transportation model, we have proposed an analytical model to calculate the drain current in ultra-thin junctionless GAA nanowire FET. The model is valid in all regions of operation, from deep depletion to accumulation and from linear to saturated regimes, as confirmed from a detailed comparison with the TCAD numerical simulations. 
The proposed model represents an essential step toward dc analysis of circuits/sensors-based on ultra-thin junctionless GAA nanowire FETs. This approach can be further used to derive a full transcapacitances \cite{6642051, JAZAERI201434} equivalent circuit and short channel effects in ultra-thin junctionless GAA nanowire FETs \cite{JAZAERI2013103}. 

\bibliographystyle{IEEEtran}

\bibliography{IEEEabrv,biblio}
\vspace{-2cm}

\end{document}